# Becoming a physicist: Major educational transition points impact women's physics self-efficacy and sense of belonging

Sarah Lindley and Chandralekha Singh

*Department of Physics & Astronomy, University of Pittsburgh, Pittsburgh, PA, United States*

skl28@pitt.edu, https://orcid.org/0000-0002-1825-9569

clsingh@pitt.edu, https://orcid.org/0000-0002-1234-5458

**Abstract:** In this investigation, we analyzed individual interviews with six female undergraduate physics majors at a large, public, research university in the US to understand their progression at different transition points to becoming physicists. Following the frameworks of standpoint theory, Schlossberg's transition theory, and domains of power, we focused our analysis on how these women initially developed an interest in physics before coming to college and how important transition points in college impacted their physics self-efficacy and sense of belonging. We find that the transitions from high school to college introductory courses and then into the physics major were bottlenecks at which women faced new challenges. Our findings suggest that although women develop initial fascinations with physics in a myriad of unique and interesting ways, they tend to follow a similar trajectory that harms their physics self-efficacy and sense of belonging over time. Finally, although we focused on women who had persisted in pursuing a physics degree thus far, their accounts point to an unsupportive physics culture that could drive other women and students from marginalized demographic groups out of the discipline.

## Introduction and Theoretical Framework

Since women have been historically marginalized in physics, improving their experiences and making their journey to becoming physicists positive is a major focus of science education research [1]. Women remain underrepresented in physics education and physics-related careers in most countries [2], including throughout all stages of college physics education [3-6]. Significant attrition of women is particularly likely to occur at specific educational transition points between major milestones, including the transition from high school to college and the transition into the physics major after completing the required first-year physics and mathematics courses [7-10]. At both points, not only does the actual course material often become more challenging, but students are also more deeply immersed into the physics culture, which has not fully recognized women as equally capable and valuable members [11-13]. Consequently, as women progress through their undergraduate physics degrees, increasing exposure to discouraging elements, such as stereotypes, can cause declines in self-efficacy and sense of belonging in physics that may push some students away from physics [14-16].

Self-efficacy, as defined by Bandura [17], refers to one's belief that they can meet a specific goal or accomplish a specific task, and, among other psychological factors, it determines what activities people choose to do, how much effort they put into them, and how they deal with related stress [18, 19]. As a result, the gender gap in self-efficacy is a major contributing factor to the underrepresentation of women in physics compared to other STEM fields [10, 20-25]. A similar problem exists with sense of belonging [26-30], another psychological factor that is correlated with self-efficacy; individuals from underrepresented backgrounds, including women, are less likely to establish strong senses of belonging in physics [31]. This lowered sense of belonging is one impact of stereotypes and biases that prevail in society, as well as in the culture of many physics departments, which can be unwelcoming for women [32-34].

In this investigation, we examine women physics majors' progression at different transition points to becoming physicists through a qualitative analysis of their experiences in physics over time. Specifically, we focus on how their self-efficacy and sense of belonging decline when transitioning from high school to college, and then again when transitioning from the first year of college into the physics major. We analyzed six interviews with female undergraduate physics majors at a large public university in the US to understand their educational trajectories, their experiences on the path to becoming physicists, and the impact of these experiences on their physics self-efficacy and sense of belonging. Declines in self-efficacy and sense of belonging for women in STEM courses have been documented in the STEM education literature in other contexts [35-38]. Our work contributes to this conversation via a qualitative analysis of individual female physics majors' experiences with physics at different critical points—from when they first developed an interest in physics, to their current engagements as undergraduate physics majors—and their impact on self-efficacy and sense of belonging over time. We find that these women, who each initially took up an interest in physics in diverse and unique ways, ultimately undergo similar negative changes in self-efficacy and sense of belonging over time, though they differ in coping strategies used to counter these changes. Their experiences indicate that in the typical physics culture and chilly climate dominated by men, in which women can feel isolated and unsupported, these transitions between educational milestones can create major roadblocks. Investigating the impact of these transitions on women who have persisted at a large research university, despite challenges throughout their journey to becoming physicists, is vitally important to create an equitable and inclusive physics culture. In physics courses, all students should be provided adequate support in an equitable and inclusive learning environment to develop a functional understanding of underlying concepts [39-47] (including laboratory courses, where gender inequities in learning have also been identified [41, 48]).

We use standpoint theory as a framework for focusing on interviews with female physics majors exclusively. Standpoint theory suggests that the viewpoints of people from marginalized or underrepresented groups are the most valuable to understand equity and inclusion issues and present solutions, since they are the people experiencing the problems at hand [49, 50]. Other individuals in physics, such as male students and professors, would not be capable of articulating the depths and nuances of the experiences of these women, since they do not know what it is like to be underrepresented in physics and may not be aware of all the struggles that students from underrepresented backgrounds face in the current physics culture. As a result, when research about this topic is undertaken from the perspectives of marginalized students rather than those in power (instructors, students from the dominant group, etc.), more objective explanations, with less interest in normalizing unbalanced power structures and maintaining social status quos, will result. This is a feature of standpoint theory that Harding calls "strong objectivity" [51]. Additionally, standpoint theory calls on researchers to be reflexive in their work, considering how social backgrounds, including their own, influence their theoretical models [52]. For these reasons, standpoint theory is our rationale for solely analyzing interviews with female physics students.

Under our broader use of standpoint theory as a theoretical framework, we also used the domains of power framework as an analytical framework. The domains of power framework is originally from Collins [53] and was used by Johnson [54] to analyze the physics culture at a small liberal arts college. In this framework, there are four domains describing how power is expressed: the interpersonal domain (between individuals), the cultural domain (in group values), the disciplinary domain (how rules are enforced and for whom), and the structural domain (the organization of power). Using this analytical framework, women's experiences that affect their self-efficacy and/or sense of belonging in physics at different transition points can be reflected back to a characteristic(s) of one or more of these domains. For instance, their experiences during their first semester college introductory physics course can be influenced by the fact that their instructors hold the positions of power in the above domains. As it relates to this work, the interpersonal domain concerns interactions between these women and their peers, instructors, family members, etc.; the cultural domain concerns the physics culture broadly; the structural domain concerns the learning environment and learning spaces (which can include the classroom, the undergraduate student lounge, etc.); and the disciplinary domain concerns whether instructors are responding appropriately to inappropriate conduct.

Using these frameworks of standpoint theory and domains of power, we analyzed the interviews to learn about women's experiences with the educational transitions into and throughout their undergraduate physics degrees. In Schlossberg's transitions model, a transition occurs when something "results in a change in assumptions about oneself and the world and thus requires a corresponding change in one's behavior and relationships" [55], providing a way to structurally examine experiences that contribute to transitions in individuals' lives from their own standpoints (e.g., [56-58]).

While standpoint theory has informed our research design as an overarching worldview (e.g., focus on women majoring in physics), we use Schlossberg's transition theory and the domains of power framework as our analytical frameworks to qualitatively examine and understand the mechanisms for the strong decline in women's self-efficacy that these transitions can cause (Fig. 1). We demonstrate that during these transitions, even women who begin their journeys extremely passionate about pursuing physics gradually start to doubt if they will ultimately succeed at physics due to the negative physics culture.

**Research Questions**

To understand their physics trajectory and impact of these transition points on the six female undergraduate interviewees, we investigated the following research questions:

1) What are some pre-college experiences that played a role in shaping these women's decisions to major in physics?
2) How did the college physics learning environment affect women's physics self-efficacy and sense of belonging as they navigated educational transitions (the high school-to-college transition, and the transition from their first-year physics courses to the rest of the major)?

We note that while there may be some overlap between research question (2) and a prior study focusing on women's experiences with male peers and instructors [46], our focus here is particularly on the transitions from high school to college and the first year of college to the rest of the major (considering women are likely to be more vulnerable at the transition points). Thus, the two papers have different research focuses; in this paper, through the two research questions, we focus on the trajectory of individual women and factors that can affect women's self-efficacy and belonging, particularly at educational transitions. Though factors such as interactions with male peers and instructors play a large role in these women's experiences, our analysis here is most concerned with how these and other factors affect these women, following individuals over time and at specific transitions, rather than a more general overview of how these particular types of experiences are part of the physics culture more broadly as in [46].

The first transition we focus on is the transition from high school to college. Societal stereotypes about who belongs in physics and who can excel in it exist through all stages of life, but these women chose to major in physics anyway, despite awareness that these stereotypes and the traditional marginalization of women in physics could produce additional challenges for them. For this to be the case, it is likely that they had some gratifying experiences or encouraging factors that helped them develop and sustain their interest in physics up to this point [19, 59, 60]. Consequently, we paid particular attention to how the women reported becoming interested in physics, and how their pre-college experiences impacted their decision to major in physics. High school plays a particularly important role, because most women decide not to major in physics before coming to college [61], and most women who do major in physics choose to do so when they are in high school [62].

The second transition of interest is from the first year of college in general to the physics major. A major contribution to the first year of college for these women is the introductory physics series. Introductory physics courses are notoriously harmful to students' self-efficacy, and to a greater extent for women than men. Over the course of these introductory courses, gender gaps in both self-efficacy and conceptual knowledge typically widen instead of close [63-66]. Sense of belonging is also an important predictor of academic success for students in these introductory courses [67]. After declaring the major, students transition into courses mainly for physics majors, which are often more mathematically challenging; spend increasing amounts of time with primarily other physics students; and at this university, gain access to a physics undergraduate student lounge. The lounge is open to all physics undergraduates and serves as a place where students can go between classes, work on assignments, or gather to study and work with their peers; it also serves as an example of a physics learning space where student behavior is not supervised or regulated by faculty members. Together, at the undergraduate level, learning spaces such as the classroom or lecture hall, laboratory, or more informal spaces like the lounge are where many of the interactions between students and faculty or between students themselves are facilitated [68-70]. These spaces contribute to establishing a learning environment that can reflect aspects of the broader physics culture, including how underrepresented students are treated and valued [61, 71-73]. When women make the transition by declaring the physics major, they become even more involved in these learning spaces, meaning that this second transition comes with greater exposure to the physics culture as it is expressed in these spaces.

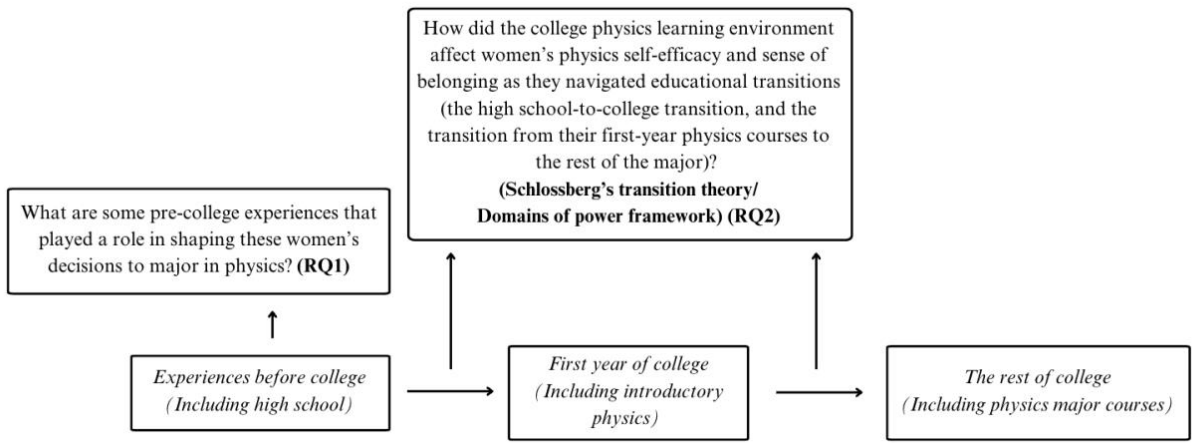

**Figure 1.** Outline of the research design. Inspired by standpoint theory, we interviewed female undergraduate physics majors discussing their trajectory, from pre-college experiences to their experiences after transitioning to college and to the rest of the major. Schlossberg's transition theory, which suggests that transitions are periods where individuals' assumptions about themselves change, and the domains of power framework, which provides a lens to examine various elements of learning environments, both served as analytical frameworks for our research questions (RQ).

## Methodology

*Interview Structure*

Over the course of one semester, 16 female undergraduate physics majors at different points in their education at a large, public research university in the United States participated in semi-structured, empathetic interviews. These interviews were in-depth, individual interviews that were each an hour long. Participants were recruited on a volunteer basis from an advertisement sent to undergraduates in the physics department and were given $25 gift cards for their participation. These interviews are considered empathetic because their goal was to gain a deeper understanding of the experiences of women majoring in physics as they navigated the physics culture in their quests to become physicists. Privacy and confidentiality were ensured by keeping the interview transcripts de-identified.

The interview protocol was semi-structured and conversational; some example questions asked in the interviews relevant for this paper are provided in Table 1. The questions are designed to be open-ended and not necessarily prompt any particular response from the interviewees. The participants gave consent for the interviews' audio to be recorded and to be quoted in academic publications. The interviews took place in person, and the audio was then transcribed through Zoom. The transcripts were then listened to again and corrected where necessary to verify the accuracy of their content.

| **Example Protocol Questions** |
| --- |
| - Can you tell me about your experiences with science, math, and physics in high school?<br>- How does college science compare to learning in high school?<br>- What contributed to your decision to major in physics?<br>- In what ways do you feel like your experiences in high school helped prepare you to study physics?<br>- How have your classes been going so far?<br>- How have your experiences studying physics changed between the first week of classes and the current time?<br>- Did you work with other people on homework, or studying for tests, in your first year in physics? What about other classes? How has that changed over the course of your studies?<br>- Do you have a sort of study group now? If so, how does it work?<br>- Can you describe an occasion where you struggled with physics?<br>- Do you have any role models/mentors? How have they supported you in physics? How has this impacted you?<br>- Do you think your experiences in physics have been different because of your gender? If so, how? |

**Table 1.** Examples of questions asked in the hour-long, semi-structured interviews.

We analyze six of these interviews here, which were selected from the 16 because they showcase a diversity of experiences from women at different stages of undergraduate education. At the time of the interviews, there were 108 total undergraduate physics majors in total, 25 of whom were women (23%, comparable to the national average); therefore, in this paper, we are analyzing interviews from 24% of the undergraduate women in the department. Among the six interviewees, two were sophomores (second-years), one was a junior (third-year), and three were seniors (fourth-years). Of note is the fact that we only interviewed students who had already declared their major, which typically happens after completing two semesters of first-year college introductory physics.

*Coding interviews*

To analyze the interviews, we utilized a combination of both deductive and inductive coding [74]. Deductive codes were based on the interview questions, which worked through the interviewees' full journey and trajectory in physics to date (e.g., high school, first year of physics, and physics major experiences). Conversely, inductive codes were decided upon during analysis based on the women's responses in the interview protocol. Our coding approach aligns with the methodological framework of Braun and Clarke's reflexive thematic analysis (RTA), which acknowledges that coders' interpretations are inherently subjective, e.g., the identity of the coders can play a role; it is a flexible approach to coding that also values collaboration between multiple coders when analyzing qualitative datasets [75, 76]. Here, after both researchers read the entire transcripts and had multiple initial discussions between them, one of the authors coded the interviews, then both authors discussed, iterated and agreed on the codes, which were inspired by our theoretical framework. The codes relevant for the findings presented here include: development of an initial interest in science/physics, and the impact of college physics learning environments at transition points in first year, and beyond first year, on sustaining, diminishing, or boosting physics self-efficacy and belonging.

Though it is easier to code responses related to development of an initial interest in science/physics based on the protocol questions, it is more challenging to code responses related to learning environments and their impact on self-efficacy and belonging at educational transitions because the interviewees are less likely to speak directly to all of these aspects at the same time. Therefore, both researchers participated in many discussions and coded iteratively following RTA [75] as described above, in order to capture subtle aspects of the women's reflections about their experiences at transitions. Using the transcripts, we identified first where the interviewees were speaking about the relevant transitions, then relevant responses related to the learning environments they experienced during and between these transitions, and lastly the nature of these environments' impacts on the women's self-efficacy and belonging. As suggested by RTA [75], it is possible for different coders to have coded these interviews differently or to disagree about the coding; however, both authors worked recursively on the coding until they agreed on all codes.

*Positionality and reflexivity*

As stated above, in RTA [75], the identities and reflexivity of researchers are important aspects that may affect their analysis. Therefore, we reflect on our identities and how they may have affected our positionality and reflexivity in conducting this research. The interviews were conducted by a woman (one of the authors), who is also a professor in the physics department. The interviewer's identity as a woman may have encouraged the female students to be more open in sharing their experiences. Both authors are women. As a result, in the context of our overarching theoretical and methodological frameworks, we acknowledge that our identity may impact our research design and analysis, including our coding and interpretations, and that other researchers with other identities may not have conducted this work in an identical way. For instance, our analysis of the interviews, as follows, reflects our own identities and experiences as women. Our identities as women mean that we may have been more easily able to understand the interviewees' standpoints and interpret them through the collaborative coding process than other individuals from different backgrounds and/or with different experiences.

In the next section, we present results related to all these codes for one individual together to help readers reflect upon each individual's physics trajectory. All names are pseudonyms.

**Results**

*Kayla*

*Development of an initial interest in science/physics*

Kayla is a sophomore astronomy major. Her interest in astronomy began with an Odyssey of the Mind program in middle school that was themed around stars. Odyssey of the Mind is an American extracurricular program in which student teams cultivate creative problem-solving skills centered around a predetermined problem and present their solutions at a competition. She shared that this program made her develop an appreciation for astronomy, but that she did not consider it as a possible career choice until she took a high school astronomy course, which solidified her interest in pursuing astronomy in college. Both of Kayla's parents have chemistry degrees and supported her interest in a STEM-based major; she is also working towards a minor in chemistry.

*High school to college transition*

Kayla went to a charter high school where every student was required to take physics, so there was an even distribution of male and female students in her high school physics class. Kayla believes that her experience with physics in high school did not provide her with a strong enough foundation for physics in college: "…I struggled first semester, even though, like, it should not have been that hard…" However, Kayla did not struggle as much with her introductory astronomy courses, even though they use similarly rigorous mathematics as the introductory physics sequence. She said that "...it was…a challenging course, but I think because it was in the context of, like, astronomy, which I thought was really interesting, like, obviously I want to learn about it," and that the introductory physics courses that taught "general physics that you need the baseline to know" was "a little bit less interesting." At the time of the interviews, Kayla was retaking the second semester of the introductory physics sequence for a better grade, so had not yet taken upper-level courses in the physics major.

When asked about scenarios in which she struggles to solve a physics problem, Kayla described her lowered self-efficacy, stating that "…college physics kind of reinforces that, like, whoa, maybe I shouldn't do physics anymore." She sometimes looks at her homework and thinks to herself: "...it's gonna be hard, and I'm not going to know how to do it," describing this as "...the feeling of dread." This, as a result, prevented her from seeing herself as a physics person: "I feel like I don't see myself as a physics person, just because…when I, like, think of, like, a physics person, like someone who's good at physics, physics comes naturally to them." Kayla originally came to the university planning to then study astronomy in graduate school but was no longer sure she would go through with this plan. She described her plan to take more chemistry classes in case she decided to change her major to chemistry.

*Impact of the college learning environment on physics self-efficacy and sense of belonging*

Kayla's overall educational trajectory shows a clear decline in physics self-efficacy and sense of belonging. She initially became interested in physics/astronomy through a very creative and hands-on avenue. Now, instead of pointing to the culture of physics and inadequate support, she feels that her struggling in college is perhaps associated with her not receiving an adequate foundation in physics throughout her secondary education. This is concerning, as introductory college physics instructors should be cognizant that their students come from different educational backgrounds and provide adequate support to all students. As Kayla was only a sophomore at the time of the interview, she was unable to share many details about her experience with the transition into the physics major, but the transition she has undergone so far has decreased her confidence in her ability to do physics. Additionally, the stereotype that physics is for brilliant and genius people [77] clearly seems to be negatively impacting Kayla, as she feels that not being able to solve physics problems without effort and struggling means she cannot be good at physics and does not see herself as a 'physics person.' She is discouraged by struggling, instead of embracing struggle as an indicator of learning. This is a cultural phenomenon that should be confronted, because believing that math and science abilities are intrinsic and not malleable is a risk factor for not persisting in the face of physics-related challenges, particularly for those from traditionally marginalized groups who do not have role models [78-80].

*Molly*

*Development of an initial interest in science/physics*

Molly is a junior physics and communication double major. She had several positive experiences with physics when she was young, beginning with reading astronomy magazines. Physics and math teachers who she shared this passion with were, according to Molly, "incredibly supportive" of her pursuing it. Additionally, she was able to meet the American celebrity physicist Neil deGrasse Tyson through her parents' place of work.

*High school to college transition*

Molly had an atypical high school experience; she went to an interdisciplinary learning high school where she was able to do independent projects related to physics. She described feeling comfortable with her teachers, because the school emphasized discussion and sharing ideas, even encouraging calling teachers by their first names, which is uncommon in the US. However, she never took exams at this high school, so her introductory college courses were her first experiences with exams. She shared that, as a result, she gets "very nervous on exams" because she "...was never taught how to study for an exam." Molly also hesitates to share her exam scores with other students in her classes and finds herself "really hesitant" to share that she is an "average student," stating that she feels "...more pressure because, part of me feels like if I don't do amazing, then it's because I'm a woman, or it reflects poorly on women…" From her perspective, because there are so few women in her courses, "...getting something wrong is kind of high stakes."

*Transition from introductory physics into the physics major*

After finishing introductory physics and moving into the rest of the physics major, Molly continued to struggle in higher-level courses, and said that her classical mechanics course was "when it started getting really, really tough, and really hard, and like, soul crushing, almost to…a point where it felt like I was working really, really hard and not seeing any results from my hard work…" She described how she "...almost gave up on the physics major…because I didn't feel like I was smart enough for it…that was my thought, was like, maybe I'm just not cut out for this…" Upon reflection, she shared how she feels male students differ from female students in this regard:

> ...every morning I wake up and convince myself that I can do it, and they don't have to give themselves the pep talk...they feel it's a struggle, but they don't think that the struggle defines whether or not they are deserving of being in the class.

Molly also decided that she did not want to pursue physics research in graduate school, expressing that her experiences had molded her to pursue politics, public policy, law, science communication, or physics outreach instead. She elaborated on her thought process behind this choice:

> …I've sort of been prioritizing the things that make me happy outside of physics more. Last year, I really prioritized physics above everything, like, I sacrificed a lot of mental and physical health because I was always studying and that really took its toll on me, so I think there's a balance that has to be achieved.

*Impact of the college learning environment on physics self-efficacy and sense of belonging*

Like Kayla, Molly had unique, creative initial experiences with physics via her interdisciplinary high school, which made her excited to pursue the discipline in college. However, over time, transitioning into college and into the physics major, she began to question her capabilities, with significant stress arising from both the introductory physics sequence and an upper-level physics class. She ultimately decided that staying in physics was not worth the toll that this constant doubt took on her mental health. Though she remained passionate about physics, it was clear that dealing with her declining self-efficacy became increasingly difficult for her over time and that she had other equally valid passions she could pursue. Even as a self-reported 'average student,' implying similar performance to most of her peers, Molly wondered whether she was 'smart enough' to pursue physics, a question that, as she suggested, male students may be less likely to ask themselves. In the current physics culture without support, many other interviewees were likewise discouraged by the impression of physics as a masculine discipline [81-83] that makes being a woman in the field more stressful.

**Mary**

*Development of an initial interest in science/physics*

Mary is a sophomore astronomy major. She had been interested in astronomy from a "very, very young age." In her junior year of high school, she took AP (Advanced Placement) Physics with a female teacher who she "loved a lot," and took AP Physics C (calculus-based) the following year with the same teacher. Mary shared that her relationship with this teacher was very strong because she "...really supported, like, us being successful physics students" and "...definitely had a huge impact on me actually enjoying physics, because she made it really easy to ask questions."

*High school to college transition*

While transitioning from high school to college, Mary took the honors version of the first semester of introductory physics in college and found that the office hours were intimidating because the professor "...would ask you to write stuff on the board in front of the whole group," and she was worried about being wrong. In her second semester, her professor would often say that certain material was "obvious," and Mary said that "the effect was...if you don't know this, you're dumb. So that made it difficult to like, ask questions in class." Mary did not feel comfortable asking questions in front of the class in either of these semesters. She also explained that the honors course had an "...air of competition with like, a lot of people need to prove that they're smart enough to be in this class, you know, either to themselves or to their peers," and that she often compared her grades to other students', fearing that other students in the class would not think she was smart enough to be there. She places significant pressure on her test grades to prove her intelligence to herself, and it is stressful for her to not know how to solve exam problems right away: "...if something, like, doesn't come out right away, I immediately, like, panic."

Mary emphasized her concerns about what other physics students think of her:

> I'm, I worried a lot about not giving my peers reason to believe that I'm smart. I felt like I had to prove it to them that I was, you know, like, smart enough to like, ask, if... needed help with homework, or like, they want to discuss a question.

She stated that she is also worried that her feminine presentation immediately discredits her in the eyes of her male peers. Mary regularly works with only one female friend, and she feels that everything would be much harder without her. Since coming to college, Mary states she has "...definitely become more insecure" about her abilities in physics. She can find it frustrating to struggle with physics, even though she knows that struggling can be beneficial to learning "...logically, but emotionally, I haven't really, like, figured it out."

*Transition from introductory physics into the physics major*

Mary started with a physics and astronomy major, but switched to just astronomy because she did not like the introductory physics courses as much as her high school courses, which she said was "...partly to do with my own anxiety, just in general, but also, it didn't really have the, like, same supportive nature." Mary was also nervous about her upcoming semester, which would be her first semester as an astronomy major instead of a physics and astronomy major, because she would no longer be in the same physics courses as her friend and would no longer be able to work with her on those classes.

Mary originally intended to have a career in astronomy research, but due to her decreased confidence in her physics abilities, is now "...thinking about being a writer and writing about astronomy, as opposed to like, actually doing the research." Though writing about astronomy for the public is still a valuable and physics-related career, Mary wondered if she would be more confident in her ability to succeed at astronomy research if her circumstances had been different: "I don't know if a more supportive environment would have made me stick to research..."

*Impact of the college learning environment on physics self-efficacy and sense of belonging*

There are several concerning aspects to Mary's progression. As she moved into more advanced courses (from high school physics to introductory physics), she felt that physics instructors and learning environments became less accepting of students coming from different backgrounds. She felt that her gender identity was also increasingly socially isolating her from her male peers and felt that her feminine characteristics prevented her from 'fitting in' in physics, a phenomenon that has been documented in other research [84, 85]. At the transition into the physics major, Mary even changed her astronomy major to no longer include physics because she was not enjoying her physics classes. A more inclusive physics culture and sources of support may have allowed her to recognise the benefits of struggling in physics learning and to fully acknowledge her own potential to be a successful physicist.

### Susie

*Development of an initial interest in science/physics*

Susie is a senior physics major. Susie knew that she wanted to pursue a science in college, but she was frustrated by her parents "pushing" her to do engineering, so she chose physics instead because she wanted to do "the hardest thing," which would be seen as "impressive." She recognized that this was "...really not a great reason to start doing physics," but felt that she developed a more genuine interest in physics after she started doing research involving condensed matter.

*High school to college transition*

Susie went to a small high school that offered one physics course, which only covered kinematics and a bit of rotational mechanics. As a result, during transition from high school to college, she felt "...not prepared for college physics with my high school background." Interestingly, her high school physics class had more female than male students. The teacher was male, and Susie thought that of the two subjects he taught, physics and chemistry, he seemed less personally interested in physics.

Once she got to college, Susie took the honors version of the first introductory physics course, which she struggled in. She continued in the honors sequence the next semester because some of the other girls in her class were doing so, and she "...was concerned that if I went to a different physics class, then I wouldn't have that support."

*Transition from introductory physics into the physics major*

Speaking of her experiences in the physics major classes broadly, Susie shared that she often felt that women were the only students struggling. She described not being as comfortable communicating answers to the class or to her male peers because she feels:

> ...nervous that if I'm wrong about most things, then people will begin to think of me as like, just not, not smart or competent...I might get the answer in the end, but if they're not there to see me get to that answer...they'll have this image stuck in their head with me just saying, like, all these wrong things. And so usually, I just don't say anything to avoid that.

Susie planned to apply to graduate programs for the following year, but had been facing doubts recently:

> ...I've recently been...on the fence if I want to apply...earlier in my, like, undergraduate career, I thought, I was like, 'yes, I'm going to go to grad school, get a PhD, and do that,' but now I'm a little nervous that that might not be for me, I can't really tell...

She was concerned about committing "...to six years in a culture…that I know I'm gonna have problems in..." Susie was working to decide "...if it's worth it, I guess, you know, to try and make things better [for women], but also, you want to be happy and succeed, too."

*Impact of the college learning environment on physics self-efficacy and sense of belonging*

Since Susie was initially determined to study the 'hardest thing' in college, she likely entered college with substantial belief in her math and science abilities; however, she clearly experienced a distinct drop in self-efficacy and sense of belonging over her undergraduate years. She became very concerned about coming off as intelligent to her male peers and did not want them to see or know about the hard work she did behind the scenes to figure things out. Like many of the other women, she felt as though not immediately being perfect at solving challenging physics problems was disheartening and even embarrassing in some capacity, a sentiment that most likely contributed to her hesitation regarding continuing physics in graduate school.

### *Melanie*

*Development of an initial interest in science/physics*

Melanie is a senior physics major. Her father was a chemist who always encouraged her to pursue science. She became interested in physics after taking her first high school physics class, sharing that she thought it might be "rewarding" to do because it was challenging. Melanie shared that the "...first principles, like very, sort of, fundamental finds" were "really appealing" and that they "...satisfy that curiosity that made me, sort of, like, bounce around so much."

*High school to college transition*

When Melanie told her high school teachers that she was going to major in physics, her physics teacher was supportive, but other teachers encouraged her to pursue a major in the humanities; however, she stuck with physics. When she arrived at college, she started getting involved with physics quickly, joining the Society of Physics Students and working on astronomy research. Throughout introductory physics, she worked in a group consisting mostly of engineering students, which initially had several women in it, but many of them dropped the course.

*Transition from introductory physics into the physics major*

After her first year, transitioning into the major, she found it more challenging to work in groups with men and experienced being talked over and getting problems explained to her without her asking.

Many female students at this university avoid the undergraduate lounge due to some inappropriate behaviors by male students that occur there. However, Melanie shared that she keeps going to the lounge "...out of spite, because I figure, like, if these people are like, terrible to women, women leaving is not going to fix that problem," even though this experience is uncomfortable for her. She described frustration at the lack of consequences for male students in the lounge who display inappropriate behaviors or language directed towards students from underrepresented backgrounds, such as women.

Due to the non-supportive and non-inclusive environment, Melanie also sometimes hesitates to go to her instructors' office hours, sharing that "...I don't want to go in there and look like…a complete idiot if I have a really, like, basic question." She generally does not seek out help from people other than a group of female friends she has that all tend to take the same courses together. Exams could also be challenging for Melanie, and they always made her nervous because "...it feels like it's a mystery every time whether I'm going to succeed on the exam" and if she hits one roadblock, she gets nervous and "...the rest of the test tends to be a disaster."

Melanie shared that she has learned to ignore inappropriate or discouraging comments from her male peers, such as those shared in the lounge, which she described as unfortunate because "...women shouldn't have to, like inherently prove themselves. It's not like women are, like, intruding on a space and have to, like, prove that they're just as good as men…" As a result, she says, when struggling with physics, "...you feel as though you're not doing well enough, and like, you're going to, like, let down your whole gender." She says that even when she does well, she tends to attribute it to some external factor, not her own knowledge, skills and hard work, which may be due to imposter syndrome, which is common in high-achieving women [86].

*Impact of the college learning environment on physics self-efficacy and sense of belonging*

Although Melanie still intends to go to graduate school for physics and finalize her career plans afterward, it was clear that college physics courses have taken a toll on her self-efficacy and sense of belonging. Initially, Melanie developed a strong conceptual interest in physics and was very passionate about pursuing the discipline, remaining steadfast in her choice to major in physics. Over time, she lost some of her self-efficacy in physics, and struggling on one problem on a test made her so anxious she would do badly on the whole exam. She developed a fear of asking her instructors and peers for help when she needed it and put immense pressure on herself to do well so she does not contribute to the cultural doubt in women's ability to excel in physics. This is reflective of stereotype threat, or the threat of being treated in accordance with a prevailing stereotype about an aspect of one's identity or of "self-fulfilling such a stereotype" [87]. Even more concerning is the fact that women who buy into societal stereotypes can experience stereotype threat to a greater extent [88].

*Melissa*

*Development of an initial interest in science/physics*

Melissa is a senior physics and astronomy major. Melissa came from a very "medically-oriented" family and for a long time felt that she would become a medical researcher. However, in her junior year of high school, she took an introductory physics course which she "just absolutely loved." From there, she enrolled in an AP physics course and an astronomy course the next year. At that point, she decided she wanted to study astrophysics. She feels that this decision shocked her family because it was so different from her previous intentions but shared that they have always been supportive of her career choices.

*High school to college transition*

Her first introductory physics course in high school was taught by a female teacher. Melissa felt that, in that course, "it [physics] kind of came naturally to me." There was a "good mix" of boys and girls in the class, and because everyone was new to the subject, she felt that the environment and atmosphere was welcoming enough for students to feel comfortable asking questions. However, in the AP course the following year, only 3 out of 30 total students were female. Melissa shared that:

> In that class, I would say that it really made me question even my ability to do physics...I think the support kind of dwindled from my teacher and even just from the people in the class, because it was all guys...every time I didn't understand something, I think I felt a little like I didn't belong there.

She feels that it was the male students contributing to this chilly atmosphere more so than the teacher, but that the teacher was unaware it was happening. However, Melissa did do well in the class overall and is glad that she did not let this course deter her from studying physics in college.

In college, Melissa formed a group of women to work with who plan to take the same classes together, especially lab courses in which they can become partners. They met each other in the spring semester of their first year and stuck together since then. During the transition to college, alongside introductory physics, Melissa also took honors introduction to astronomy with one other female student. Through the introductory physics courses, she learned that people pretend to understand more than they do and that it is only when she leaves the room that she realizes that this does not reflect reality and that everyone was trying to figure the material out.

*Transition from introductory physics into the physics major*

Melissa did research in physics her first two years and then switched to research in aerospace engineering at a neighbouring university. Within one year of being on a team for an aerospace engineering project, she was able to progress to lead systems engineer. Because she greatly enjoyed this experience in a supportive environment, unlike her experiences in physics, she switched her plans and now is planning to pursue graduate studies not in physics, but in aerospace engineering. She indicated that there were deep cultural problems in the physics department she was tired of that she did not experience in her time working on aerospace engineering, e.g., sharing that:

> ...our undergraduate student lounge...is not somewhere that I go anymore...the discussions that have happened in there and the general atmosphere from the male students is offensive...not something that I want to be working and hearing at the same time. Or ever.

She elaborated that inappropriate behaviours have been directed both at herself and at other female students, and that some male students told her that those conversations do not stop when the women speak up, they just continue after the women leave. Melissa reflected that "...that's just something I have experienced largely from physics...that I actually haven't experienced in my work at [neighbouring university]."

When discussing her thought process when struggling on a physics problem, Melissa stated:

> I think, like, okay, what's wrong with me? Why can I not do this? Or like, I should know this and I'm not right now, and I just feel dumb. You know what I mean? I'm usually very hard on myself.

These feelings give her significant test anxiety when she struggles with problems on exams. She added that, though she is comfortable asking her group of female friends for help, she would not ask the male students because of past experiences with male peers acting surprised if she does not know something. She added that those types of incidents make her experience self-doubt, sharing, "...it's weakening to me to feel like I'm struggling on something...it scares me to think that...I'm not going to be successful."

Melissa emphasized the differences she experienced between the physics culture and aerospace engineering culture:

> I'm not afraid to ask questions, and I know that I'm new to the material...I know what to do now when I need help or need to ask questions...I'm not as hard on myself. Somehow, still, though, with physics, like, my current courses, I'm still nearly about that hard on myself...in engineering...I was the only woman on my entire team this summer, but I was the leader and it actually shows that I was valued and respected.

Additionally, in physics, Melissa has "...never been told directly...'you're doing a good job'...I never really had the reinforcement from a professor or TA."

*Impact of the college learning environment on physics self-efficacy and sense of belonging*

Though Melissa is clearly still enthusiastic about pursuing a STEM career more broadly, it is evident that the physics culture specifically had a substantially negative impact on her self-efficacy and sense of belonging in physics. One need only compare her excitement when describing her aerospace engineering experience to the dejection she expressed about struggling in physics, interacting with male peers and lack of positive recognition from instructors to see why she may have felt that aerospace engineering was a better option for her.

## Discussion

### Role of pre-college experiences in shaping decisions to major in physics

Among the women, there are some general trends in factors contributing to first developing an interest in physics, such as being exposed to it through popular media like books and documentaries. Many of the women also had parents who supported and encouraged their interests in physics, especially parents who have careers or degrees in STEM. Previous research has also found links between parents' encouragement and attitudes about science and students' interest in pursuing a science-based career [89-93].

The women had variable experiences with physics in high school, and there may be a connection between the quality of high school physics instruction and perceived preparedness for college introductory physics. High school can be crucial for enhancing women's interest in physics and building up the skills and resilience to persist despite future challenges in this male dominated field with a chilly climate, as discussed here. However, representation in high school physics at the AP level is already stratified [94]. Additionally, previous research has found that women often report receiving more recognition from their high school physics instructors than from their college instructors [95], similar to the case for many women in our study.

We find that key factors contributing to these women gaining an initial interest in physics included being exposed to physics through popular media or related educational programming (e.g., Kayla, Molly), being exposed to and encouraged to engage with science by parents (Kayla, Molly, Susie, Melanie), and having positive experiences with physics in high school (Molly, Mary, Melanie, Melissa), including positive relationships with high school physics teachers. However, Melissa described negative experiences with her advanced placement physics course, and several women felt that their high school physics courses did not provide them with strong enough foundations for college introductory physics (Kayla, Molly, Susie). Considering these findings through the domains of power framework [54], it appears that while Melissa experienced problems with the interpersonal and cultural domains in high school physics (e.g., she felt unsupported in AP physics because she was significantly outnumbered by men in her class), the characteristics of all four domains were at least not unwelcoming enough to deter these women from choosing to major in physics in college.

### *First transition: High school to college*

Moving into the first year of college, statements from Kayla, Molly, Mary, and Susie suggest declines in physics self-efficacy as a result of their introductory classes. Key factors contributing to this decline included feeling that their high school physics foundations were not strong enough (Kayla, Molly, Susie), feeling additional pressure to succeed due to their gender (Molly, Susie), feeling intimidated by instructional methods that put one on the spot (Mary), and the competitive nature of the honors introductory course in particular (Mary). At this stage, it seems that the women struggled the most with the interpersonal and cultural domains of college physics, as they are more exposed to the culture of physics, which devalues struggling and emphasizes competition; this culture also presents itself through their interpersonal interactions with their male peers.

The first transition, into the introductory physics courses, was a source of significant stress for many of these women, who worried about being underprepared from high school when they struggled with college physics, were concerned about losing the limited support they had in the form of other female classmates, faced test anxiety, and hesitated to seek help from their instructors due to lack of perceived recognition and support in the physics culture. Prior research corroborates that these courses can often be harmful to the self-efficacy of all students, but even more so for women [96]; additionally, on average, men majoring in physics often tend to start these introductory physics courses with more prior preparation than their female peers and without adequate support for underrepresented students, men are, on average, still ahead when the introductory sequence is over [97, 98]. Thus, it is critical to reflect upon the findings presented here about how these women were negatively impacted by the physics culture and how physics instructors or male peers did not support them in the existing physics culture, to make the physics learning environments equitable, supportive, and inclusive.

*Second transition: First year of college to the physics major*

As the women progressed into the rest of the physics major, they continued to experience declines in their self-efficacy and sense of belonging in physics. Kayla, Molly, and Melissa all shared that they can get frustrated when they are not able to immediately solve a physics problem. Kayla, for example, said that she doesn't see herself as a "physics person" because physics does not come as naturally to her. Melanie and Melissa both shared experiences with test anxiety. Most concerningly, Kayla, Molly, Mary, Susie, and Melissa (five of the six women) all either no longer plan to pursue a career in physics or astronomy research and/or are having doubts about going to graduate school for physics or astronomy after their experiences with undergraduate physics. Only Melanie is still confident in her plan to study physics in graduate school.

For these women, it seems that the physics culture is a primary driver of this problem. Consistent with the domains of power framework, the idea in the physics culture that one needs to be brilliant and naturally good at physics [77-80] is harmful. For instance, Molly shared that she feels her struggling defines whether or not she deserves to be in her physics classes, and Melissa berates herself for struggling with problems. Mary shared that she has become insecure about her abilities in physics. Molly, Mary, Susie, and Melanie all said that they feel a need to 'prove themselves' to their male peers, though Melanie stated that she knows she shouldn't have to do this. Mary thinks that her femininity presents an additional barrier to overcome to prove herself, reflecting the stereotype of physics as masculine [81-85]. These results indicate that these women are experiencing stereotype threat [87, 88] while trying to prove they can fit in with the image of a physicist.

Additionally, these women faced problems while interacting with male peers and instructors. Melissa and Melanie both described the uncomfortable and inappropriate environment established in the undergraduate lounge. Melissa described feeling judged in some cases when she asked male students for help on something she didn't know. Melanie shared that she feels it is hard to work with male peers without being talked over and finds that instructors' office hours are not supportive environments to ask questions that seem too basic. All of the above reflect issues can be understood within the domains of power framework. For instance, these women experience problems with their interpersonal relationships with their male peers and professors (interpersonal domain). There is also an issue with the disciplinary and structural domains, with male students being allowed to perpetuate an unwelcoming and 'offensive' environment in the undergraduate lounge.

After transitioning to the major, in an unsupportive physics culture, most of the women reported feeling discouraged, at least at some point, e.g., when struggling on physics problems. They shared a sense of frustration when they could not solve physics problems effortlessly and were fearful or hesitant to ask male peers and instructors for help. The women were aware going into the physics major that physics would be challenging, and still wanted to pursue it anyway; in particular, Melanie and Susie chose physics in part for this reason. However, in a culture where struggling is seen as a negative thing, finding physics to be difficult could be more frustrating for these students when they already felt discredited due to their gender. In an unsupportive physics culture, these frustrations may partly be related to stereotype threat [87] and the pressure on women to excel in physics to represent their demographic well, as Molly, Melanie, and Melissa described. We find that without appropriate support, in the inequitable college physics culture, their gender also affected their physics sense of belonging and was closely tied to their self-efficacy, consistent with prior studies [99-103].

*Conclusion*

Our interviews show general trends that outline overall similarities of the interviewees' trajectories in physics. Most became fascinated with physics when they were young through a variety of avenues, appreciated both conceptual and practical aspects of physics, had families who encouraged them to pursue this interest, and were very excited to do so in college. However, as they proceeded through their education, they gradually developed more and more fear of not being able to succeed in physics, with this doubt exacerbating during introductory college physics for some and worsening when transitioning to upper-level coursework. In Schlossberg's transition theory, individuals' responses to transitions are affected by many factors, including their perception of the transition itself, characteristics of the individual undergoing the transition, and their social environments before and after the transition [55]. From this, one can see how learning environments and cultures can affect individuals undergoing educational transitions. In this case, when women are undergoing transitions related to their physics education, and therefore are more particularly vulnerable to changes in how they feel about themselves as physicists or "physics people," the physics learning environment can play an important role. Our interviews suggest that the unsupportive physics culture contributes to the major finding of our analysis, which is that the women's changes in self-efficacy and sense of belonging related to these transitions are largely negative. Thus, our research provides a possible mechanism to make sense of prior quantitative research which shows that women drop out of many STEM disciplines, including physics, with significantly higher GPAs than men [9]. The women we interviewed were able to persist in physics perhaps more than one might expect based on their negative experiences, potentially by developing support systems, such as Susie and Melissa working with other groups of women, while other women who do not have such support systems may struggle even more to remain in the major. The fact that many of the women interviewed had parents with careers in science may also have contributed to their persistence in physics. These are both support systems that could potentially help individuals undergoing transitions to adapt to the transitions, according to Schlossberg [55]. Still, even with these sources of support, most of the women doubted whether they will continue to do research or study physics in graduate school.

Our interview findings show that although these women have persisted in physics up until this point, the physics learning environment in this department is not adequately equitable and inclusive, and that the unsupportive culture of physics has the potential to negatively impact women's physics self-efficacy and sense of belonging on their paths to becoming physicists. Melanie and Melissa, for example, described inappropriate behaviors by male peers in the undergraduate lounge; multiple women described being talked over or explained to by their male peers, being afraid to ask professors and peers for help when it was needed, and feeling increased pressure to do well in physics because they were women. As most of these problems either first arose or worsened at both post-high school transitions, it is logical to conclude that the college physics setting and the physics culture is a significant contributor to these declines in sense of belonging and self-efficacy. Such trends reflect failure of the physics department in various domains in Johnson's domains-of-power framework [54], e.g., women's negative interactions with their peers and instructors (interpersonal domain) are fueled by stereotypes about physics (cultural domain) in an environment that is designed neither to support them educationally (structural domain) nor to intervene in their favor (disciplining domain). As students progress through their education and become more knowledgeable in their discipline by undergoing various transitions, they should ideally experience an increase in self-efficacy and sense of belonging, not a decline. The women including women of color in Johnson's work, at a small college, follow this positive trend [54]; however, our research here shows the reverse trend among women at a large research university. This university functions more like a prototypical physics department across these domains, which is harmful for women particularly at these transitions, where they are more vulnerable to changes in their self-image and feelings about their abilities in physics.

What are the underlying roots of this phenomenon? Physics departments that enable inequitable and non-inclusive physics learning environments, reflecting weaknesses across all four domains, can be harmful for traditionally marginalized students such as women. Professors and male peers directing micro- or macroaggressive behaviors towards female students may or may not even be aware of the impact of their behaviors [104], and in departments with cultures like the one at this university, nobody observing is likely to intervene in these situations. For instance, once these women transition into their major, they witness the behaviors in the undergraduate lounge that are so inappropriate that they have driven most of these women away from it altogether. As the lounge is a space dedicated solely to physics majors, this exodus and lack of departmental action is representative of a failure in the disciplinary domain [54] and reflects how these women are not adequately valued or respected in the physics department. Our interview findings, which suggest that such tendencies become increasingly prevalent as women progress through their major and become more immersed in the physics culture, could potentially explain why quantitative studies show women leaving the discipline with higher grades than men who leave [9]. Without support in an inequitable physics culture, at each transition, we find that women's self-efficacy and sense of belonging suffers driven by cultural conditions that disserve them.

Physics departments should focus on improving their culture and making the learning environments equitable and inclusive to ensure closure of the gender gaps in self-efficacy and sense of belonging in physics [10, 21, 26, 29-31, 35-38, 105, 106]. To revisit more broadly the idea of standpoint theory [49], one of the best things we can do is continue to listen to female undergraduate physics majors when they share their experiences like the ones presented here. They are traditionally marginalized students living these experiences, and they know best how challenging it is to stay afloat in an inequitable and non-inclusive physics culture. Listening to and believing in these women when they describe their experiences or their suggestions is the first step towards reimagining the physics culture in which they are seen, heard, and positively recognized as equally capable physicists as men. For structural or systemic changes to be made in how instructors, peers, and other physics department members interact with and treat students from underrepresented backgrounds, instructors must be cognizant of a few things. First, they must recognize that there is a problem with the status quo in the physics culture, consistent with how the domains of power operate in a prototypical physics department like the one in this study, and second, they must recognize that by staying neutral [107], they are letting the culture perpetuate. If instructors do not try to understand the viewpoints of female students, they are unlikely to realize that certain behaviors in their office hours (trivializing difficulties, questioning students' knowledge, etc.; see [108]), regardless of the intentions behind them, can make women anxious, decrease their self-efficacy, and make them hesitant to seek out help in the future. Standpoint theory, beyond an academic concept serving to rationalize interviewing these women, is a necessary perspective for physics departments to take and put themselves in the shoes of traditionally marginalized students. Seeing positive change in this regard is not a hypothetical or unattainable goal, as its success has been proven at a small US institution [54]; it simply needs to be scaled adequately and implemented enthusiastically, keeping in mind all domains of power and how they interact with each other [54]. For systemic change in the physics culture, instructors must develop growth mindsets about the potential of all their students [109], especially bearing in mind the impact that perceived recognition by instructors can have on women's self-efficacy [20, 23].

Following from this, and the results of other qualitative interviews regarding this issue (i.e., [1]), there are some tangible changes that would be beneficial for physics instructors and departments to enact to meet this goal. Those in the position of power, such as instructors and department chairs, should work to improve the physics culture along all domains of power. For example, improving along cultural (establishing that physics is collaborative and learning physics involves working hard using deliberate approaches), interpersonal (instructors and peers are friendly and helpful instead of being rude and perpetrating microaggressions) and structural domains (e.g., improving physics learning environments including classrooms, student lounge norms etc.) can go a long way. Institutionally, physics departments should make clear to instructors their responsibility to intervene when they witness inappropriate or exclusionary behavior to strengthen the disciplinary domain. Along this vein, they should make strong efforts to establish community amongst students, for instance, by establishing student groups and emphasizing the importance of collaboration in the classroom, which will contribute to strengthening the interpersonal domain.

Instructors should also be mindful of the learning environment established in their classrooms and the physics cultural domain more broadly. For instance, knowing the self-efficacy gap found between their male and female students, they should make efforts to dismantle any fixed mindset they may have about students' possible achievements [109]. They should recognize that students from marginalized backgrounds, like those from dominant groups, can become successful physicists if instructors believe in them, recognize them, and provide appropriate support to help all students succeed. Lastly, instead of trivializing information and demoralizing students when they are struggling, instructors should aim to serve as both an academic resource and a source of support in these cases, emphasizing the normalcy and benefits of struggling [110]. If departments enthusiastically attempt to implement such changes, the benefits to the physics culture and learning environments will help bolster marginalized students' ability to adapt to educational transitions before and throughout their undergraduate years, while simultaneously minimizing any decline in self-efficacy and sense of belonging they face during these transitions.

**Limitations and future directions**

Though the results of these interviews are insightful, there are several limitations to this work. The sample size of women interviewed is small, and the responses of the women interviewed may not perfectly encapsulate the viewpoints of all women in this physics department. Additionally, many of the women interviewed come from more privileged academic backgrounds and had parents who were scientists, which may have contributed to their persistence in physics. This, alongside with the sources of support many of these women were able to find throughout their undergraduate years (e.g., making friends with other women in the department), most likely contributed to the women remaining in physics, even when facing an unwelcoming and chilly physics culture. However, this does not directly explain why other women leave. Future work should focus on comparing the trajectories and outcomes of women physics students who do and do not have such networks of support before and during their undergraduate years, to further build on our understanding of attrition of female physics majors. Additionally, other qualitative methods besides interviews could be used to explore this issue, such as surveys or reflective journals [57, 111], etc.

Additionally, this research focuses specifically on undergraduate female physics majors at a large, public research university in the US. Other types of institutions from around the world should be considered in future research, including smaller liberal arts colleges, and minority-serving or women-only institutions. Moreover, future interviews should focus on the impact of other aspects of students' identities, as well as the intersectionality of different identities (e.g., gender and race) on students' experiences in physics [112-114].

**Ethics statement**

This research was carried out in accordance with the principles outlined in the University of Pittsburgh Institutional Review Board (IRB) ethical policy, the Declaration of Helsinki, and local statutory requirements. Informed consent was obtained from each interviewed student to participate in the study and for their comments to be published.

**Data availability statement**

Any data that supports the findings of this study are included within the article.